\documentclass[pre,showpacs,notitlepage,twocolumn,superscriptaddress]{revtex4-1}

\pdfoutput=1
\usepackage{tabularx}
\usepackage{blindtext}
\usepackage{svg}
\usepackage{float}
\usepackage{mwe}
\usepackage{multirow}
\usepackage{hyperref}
\hypersetup{colorlinks,linkcolor={blue},citecolor={blue},urlcolor={blue}}

\usepackage{graphicx}

\usepackage{bm,amsmath}

\usepackage{latexsym}

\usepackage{verbatim}

\usepackage{xcolor}
\usepackage{subfigure}

\usepackage{gensymb}

\usepackage{soul}

\usepackage{tikz}

\newcommand{\sara}[1]{{\color{red} #1}}

\renewcommand*\vec[1]{\mathbf{\bm{#1}}}
\newcommand{\revise}[1]{{\color{black} #1}}
\begin{document}

 \title[]{Conformation and dynamics of wet tangentially-driven active filaments}

\author{Loek van Steijn}
 \altaffiliation[]{These authors contributed equally.}
\affiliation{Institute of Physics,  University of Amsterdam, Amsterdam, The Netherlands
}%
\author{Mohammad Fazelzadeh}
 \altaffiliation[]{These authors contributed equally.}
\affiliation{Institute of Physics,  University of Amsterdam, Amsterdam, The Netherlands
}%

\author{Sara Jabbari-Farouji}
 \altaffiliation[Correspondence to: ]{\hyperlink{s.jabbarifarouji@uva.nl}{s.jabbarifarouji@uva.nl}}
\affiliation{Institute of Physics,  University of Amsterdam, Amsterdam, The Netherlands
}%
 \email{Correspondence to: s.jabbarifarouji@uva.nl}


 \date{\today}

  \begin{abstract}
We explore the impact of hydrodynamic interactions  on the conformational and dynamical properties of wet tangentially-driven active polymers using multiparticle collision dynamics simulations. By analyzing active filaments with varying degrees of flexibility, we find that fluid-mediated interactions significantly influence both their conformation and dynamics. These interactions cause polymer conformations to shrink \revise {relative to their dry counterparts}, especially for semiflexible polymers at high activity levels, where the average size of wet chains becomes nearly three times smaller, due to \revise{local buckling of wet polymers}. This hydrodynamic-induced shrinkage is a hallmark of active polymers, as fluid-mediated interactions \revise{do not affect conformational properties} of passive polymers. Furthermore, for tangentially-driven polymers where activity and conformation are coupled, hydrodynamic interactions significantly enhance the orientational and translational dynamics compared to their dry counterparts.


  \end{abstract}


\maketitle
 \section{Introduction} 
 
Active particles, driven by self-generated mechanisms, have emerged as a fascinating  research area as they exhibit novel  patterns of  non-equilibrium self-organization with no counterparts in equilibrium systems. Active systems, encompassing a diverse range of entities such as bird flocks, fish schools, swimming bacteria, and artificial microswimmers, offer valuable insights into the  governing principles of non-equilibrium self-organization in active matter~\cite{vicsek2012collective,bechinger2016active}. Recent research interests have extended beyond point-like or rigid active particles (e.g., active Brownian particles) to include particles with complex shapes or internal degrees of freedom. Prominent examples include active polymers and filaments~\cite{active_poly_review,active_pol_20,stark-rigid-pointlike,gomper-review-pointlike-rigid}, which have recently attracted a lot of attention.

Active filaments and polymers can be found on a wide range of scales from ATP-powered chromatin in cell nucleus and  motor-driven actin and microtubules to microscopic worms and snakes. Inspired by experiments of active biological filaments a diverse range of theoretical and computational models for active polymers have been proposed in recent years~\cite{active_poly_review,active_pol_20} including scalar~\cite{kremer,LikosKremer} and force-driven active polymer models where the active force can be  stochastic~\cite{ABPO_ANALYTICS-eisenstecken_gompper_winkler_2016,GHOSHABPO}, tangential~\cite{Isele,Active_flexible,dreyfus2005microscopic-tangential,PhysRevResearch.6.L032002-tangential} or perpendicular~\cite{GolestanianTransversleyActivated} to the backbone. Among these, the tangentially-driven active polymer model has garnered increased interest due to its relevance to filament-like systems, such as motor-driven actin and microtubule filaments, elongated bacteria like Proteus mirabilis~\cite{lough_2023}, and  worm species such as \emph{T. tubifex}  ~\cite{aquarium_worms} that propel themselves along their backbone.


Microscopic active filament-like systems, such as motor-driven biopolymers and bacteria swim in aquatic environments. Consequently, their shape and dynamics are influenced not only by their intrinsic activity mechanisms but also by fluid-mediated interactions. It is well known that the dynamics of passive polymer solutions are significantly affected by hydrodynamic interactions \cite{Zimm-Model, doi-edwards,passive_wet_polymer,Tylor-mpcd}. Therefore, it is important to elucidate the interplay between internal degrees of freedom, activity, and hydrodynamic interactions on the conformation and dynamics of active polymers. Integrating shape, activity, and hydrodynamic interactions in studies of active polymer models not only provides a deeper understanding of biophysical processes involving wet active polymers but also inspires the design of artificial filamentous microswimmers with tailored properties.

The majority of studies on active polymers have ignored hydrodynamic interactions and have focused on the interplay between internal dynamics and activity mechanisms. There are a few studies that investigate the role of hydrodynamic interactions on the structural and dynamical properties of active filaments~\cite{lough_2023,Barriuso_Reviewer1-DPD,tylorYeomansHelical,stark-zottl-yeomans,Adhikari_2012, laskar2015brownian, Canio_2017, saintillan2018extensile, martin2019active, ABPO-N200, santillan-strech-coil, Swilrling-Goldstein,jiang-active-Filaments-confined-OseenTensor}. These  include studies of active Brownian polymers in solution~\cite{martin2019active, ABPO-N200} and active polymer models where monomers are subject to permanent or stochastic active force dipoles along their backbone~\cite{Adhikari_2012, santillan-strech-coil}. These investigations reveal the significance of hydrodynamic interactions on polymer dynamics and even on conformational properties.

In this work, we analyze the conformational and dynamical properties of tangentially-driven active polymers~\cite{Isele}  embedded in a fluid where the monomers are externally actuated, focusing on  unconfined dilute active polymer solutions. The dry version of this model has been subject of intensive studies~\cite{Isele,Fazelzadeh,Active_flexible,Mokhtari,inertia-spiral,khalilian2024structural,collective-Gompper,Prathyusha,partially-Emanuele}. 

 \revise{Biological active filaments driven by molecular motors, as well as chains of Janus active colloids bonded through electrostatic or magnetic dipolar interactions~\cite{Active_chains_Granick,sakai2020active_dipolar_colloids,traveling_string}, are examples of systems where the active force is constrained to be tangential to the backbone. These systems are force-neutral and, at the hydrodynamic level, can be described as chains of hydrodynamic dipoles aligned tangentially to the backbone, similar to the models developed in~\cite{Adhikari_2012, santillan-strech-coil}. In contrast, active polymers in motility assays driven by molecular motors fixed to the assay are not force-neutral and they can be considered as externally actuated~\cite{MicrotubulesCounterclockwise-assay,keya2020synchronous-assay}. Moreover, externally-actuated active polymers with tangential driving can be engineered by applying a locally tangential drive to a chain of colloids with flexible linkers using dynamic optical tweezeing~\cite{CURTIS2002169-dyntweez,LyngeEriksen:02-dyntweez,ghosh2019all-dynTweez}.}
 
Comparing the conformation and dynamics of dry and wet tangentially-driven   polymers  we address two important questions.  (i) Do the  hydrodynamic interactions  lead to the emergence of new modes of conformation and transport of active polymers? (ii) How do the longtime dynamics of the wet active polymers differ from that of the dry ones?

To account for hydrodynamic interactions, we employ the explicit fluid simulation method of multi particle collision dynamics (MPCD). The MPCD is a mesoscale particle-based simulation method for simulation of fluids, which naturally incorporates thermal fluctuations. The MPCD is able to capture the hydrodynamic correlations correctly~\cite{Malevanets,Arash-review,mpcd-Gompper,Llahi,HI-Correlation-MPCD} and is widely used for simulating dynamics of colloidal and polymeric solutions~\cite{Tylor-mpcd, Malevanets,tylorYeomanMPCDNematic,efficientGPUMPCDMPCD,Lamura-MPCDExampleMPCDMPCD,lamura-attractivePolMPCDMPCD,bolintineanu2014particleMPCDMPCD,HScreening-RandomMPCD-NoHIMPCDMPCD,ZaccarelliMPCDMPCD,ArashMPCDMPCD-sedimentation,hecht2005simulationMPCDMPCD,padding2006hydrodynamicMPCDMPCD,gompper2009multiMPCDMPCD}. The MPCD method allows to faithfully reproduce the fluid-mediated interactions with significantly less computational costs. 
In the context of active systems, it has been utilized to study the dynamics of microswimmers such as generic squirmer model~\cite{Stark-swimmer-MPCD-africanTrypanosome,eColi-mpcd,squirmer-stark,squirmer-gotze,zottle-mpcd} and wet active Brownian polymer models~\cite{martin2019active,ABPO-N200}.  In this article, we present an implementation of wet tangentially-driven polymers in MPCD. The polymer model consists of linear bead-spring chains where each monomer experiences an external active force tangential to the local tangent of the backbone~\cite{Isele}.

 We investigate the conformational and dynamical features of wet tangentially-driven polymers with varying activity levels and two different degrees of flexibility, comparing them with those of their dry counterparts. Our investigations reveal that fluid-mediated interactions influence both the conformational and dynamical properties remarkably. Hydrodynamic interactions cause polymer conformations to shrink, an effect that becomes more pronounced with increasing activity levels and bending stiffness.  At high activity levels, the conformations of semiflexible polymers 
\revise{buckle and locally fold} due to   consecutive extensional and compressional flow profiles created  by active forces around  the polymer backbone which compete with its bending elasticity. \revise{The conformations of wet semiflexible active polymers resemble qualitatively to buckled conformations of microtubules in motility assays~\cite{MicrotubulesCounterclockwise-assay,keya2020synchronous-assay}, which form rings and spirals.}
 Wet active polymers exhibit faster orientational dynamics and enhanced longtime diffusion coefficient. Interestingly, the orientational relaxation time of the end-to-end vector of the polymers is almost independent of their degree of flexibility but decreases with increasing active force.

The remainder of this article is organized as follows. In section \ref{sec:simulation}, we first introduce our simulation setup  including the implementation of tangentially actuated polymer model in the MPCD, providing definition and values of the relevant dimensionless parameters. In Sec.~\ref{Sec:conformation}, We investigate the effects of hydrodynamic interactions (HI) on conformational properties of active polymers. In Sec.~\ref{Sec:dynamics},  we  discuss the shrinkage mechanism induced by fluid-mediated interactions. Then, we present the results on impact of hydrodynamic interactions on orientational and translational dynamics of active polymers.
We end our paper with a summary of our most important findings and  concluding remarks in Sec.~\ref{Sec:conclusion}.

 \section{Simulation details \label{sec:simulation}}
 
\subsection{Multiparticle Collision Dynamics simulations}
 In the MCPD approach, the fluid is simulated by many  identical point-like particles, which undergo discrete iterative streaming and collision steps. During each step, the MCPD particles propagate and locally interact with each other \cite{Malevanets}. In the streaming step the fluid particles of mass $m$ are propagated by solving Newton's equations of motion over a small  time step $\delta t$ using the Velocity-Verlet integration scheme given by
\begin{equation}
\begin{split}
    \vec{v}_i(t + \frac{1}{2}\delta t) = \vec{v}'_i(t) + \frac{\vec{f}_i}{2m}\delta t \\
    \vec{r}_i(t+\delta t) = \vec{r}_i + \delta t \vec{v}_i(t + \frac{1}{2}\delta t) \\
    \vec{v}_i(t + \delta t) = \vec{v}_i(t + \frac{1}{2}\delta t) + \frac{\vec{f}_i}{2m}\delta t
\end{split}
\label{eq:Velocity-Verlet}
\end{equation}
Here, $v_{i}$ and $r_{i}$ are the velocity and position of monomer $i$ respectively,  $v'_i$ is its updated velocity vector  after multiparticle collision step given by Eq. \eqref{eq: SRD}, and  $\vec{f}_i$ is the force acting on the particle $i$.
After streaming, the solvent undergoes a multiparticle collision representing the core of the MCPD algorithm.
To implement the collision steps  the fluid particles are sorted into cubic cells of edge length 
 $L_{cell}$, which sets  the spatial resolution of HI. Thereafter, the relative velocity of each particle with respect to the centre of mass velocity of all the particles in the cell is rotated by some constant angle $\alpha$ about a randomly selected axis. This collision method is called Stochastic Rotational Dynamics (SRD) \cite{Malevanets,Ihle2001}. \revise{The choice of collision rule and its parameters  determines the transport coefficients of the mesoscale MPCD fluid~\cite{MPCD_viscosity}.} Update of velocities of all particles in the same cell \revise{with $N_c$ MPCD particles}  undergoing a stochastic collision is described by  

 \begin{align}
\vec{v}'_i (t+\delta t)  &= \vec{v}_{cm}(t+\delta t)\nonumber \\
&+ \mathbf{\mathcal{R}}(\alpha)\cdot [\vec{v}_i(t+\delta t) - \vec{v}_{cm}(t+ \delta t)],
    \label{eq: SRD}     
 \end{align}
 where $\mathbf{\mathcal{R}}(\alpha)$ is a \revise{stochastic rotation matrix with angle $\alpha$  and $\vec{v}_{cm} = \frac{1}{N_c}\sum^{N_c}_{i=1}\vec{v}_i$ is the centre of mass velocity of a collision cell.}
The updated velocities $\vec{v}'_i (t+\delta t)$ are then used for  repeating the streaming step described by Eq.~\ref{eq:Velocity-Verlet}. \revise{The SRD method is not isothermal on its own as it naturally imparts the NVE ensemble to the system. 
 To apply a thermostat with constant temperature $T$, we use a Maxwell-Boltzmann-scaling (MBS) approach~\cite{Thermostat-viscosity}.  This thermostat rescales the  relative velocities $\Delta\vec{v}_i=\vec{v}_i-\vec{v}_{cm}$  of the particles in a collision 
cell by a constant factor $\mu$ after each collision step using Maxwell-Boltzmann distribution of kinetic energy.  The rescaling factor $\mu$ typically differs for every cell and collision-time step. Since the total
relative momentum of a collision cell is zero, such a scaling leaves the total momentum of a cell intact.
The MBS has been proven to render the correct hydrodynamics properties of isothermal fluids~\cite{HUANG2010168}. 
To obtain $\mu$ for each collision cell, at every collision, a random kinetic
energy  $E_{k}^{'}$ is taken from the distribution function  of kinetic energy of the MPCD ideal-gas particles given by Maxwell-Boltzmann distribution~\cite{HUANG2010168,Thermostat-viscosity}:
\begin{equation}
    P(E_k^{'}) = \frac{1}{E_k^{'}\Gamma(\frac{f}{2})}\Bigg(\frac{E_k^{'}}{k_BT}\Bigg)^{\frac{f}{2}}\exp \Bigg(-\frac{E_k^{'}}{k_BT}\Bigg).
    \label{eq:MBS-dist}
\end{equation}
Here, $\Gamma$ denotes the Gamma function and $f=3N_c-1$ is the number of the degrees of freedom of the cell and the mean kinetic energy is given by $\langle E_k\rangle= f/2 k_BT$.

Then, the   respective scale factor for the velocities is obtained as
\begin{equation}
 \mu = \Bigg(\frac{ E_{k}^{'}}{E_k}\Bigg)^\frac{1}{2},
    \label{eq:MBS-scaling}
\end{equation}
where $E_k=m/2 \sum_{i=1}^{N_c}\Delta\vec{v}_i^2$ is the kinetic energy in the center of mass  frame of each cell before rescaling.
}

 \begin{figure}[t]
    \centering
    \begin{tikzpicture}
            \draw (0,0) node[inner sep=0]{\includegraphics[width=1\linewidth ]{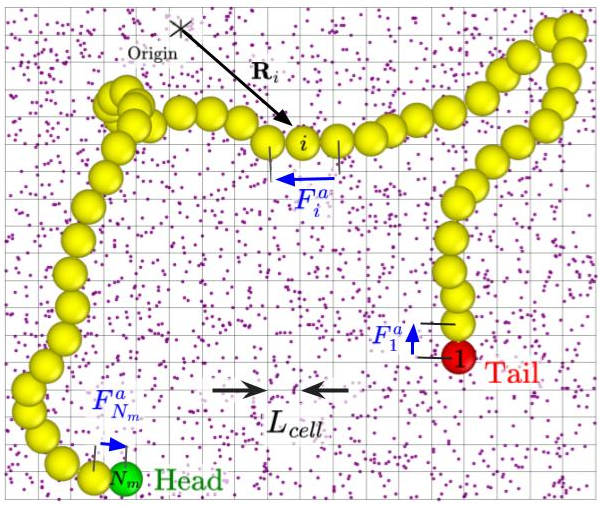}};
    \end{tikzpicture}
    \caption{Schematics of the active tangentially driven polymer,MCPD  particles andMCPD  mesh. The polar nature of this particular type of activity let us define a head and a tail for the filament.}
    \label{fig:Schematic}
\end{figure}
 
 \subsection{Active polymer model}
 
 We model a semiflexible polymers  by a bead-spring  chain composed of $N_m$ active monomers of mass $M$, with positions $\vec{R}_i$, and velocities $\vec{V}_i$ $(i = 1, . . . , N_m )$, see Fig.~\ref{fig:Schematic}. The monomers are connected by harmonic springs described by potential $U_{sp}$. The degree of flexibility of polymers is incorporated
  by restrictions of bond orientations via a bending potential $U_{bend}$. The self-avoidance is ensured via excluded-volume interactions $U_{EV}$.
The dynamics of each monomer is described by the following equation of motion
 \begin{equation}
         M \frac{\partial^2}{\partial t^2}\vec{R}_i = -\nabla  U_i + \vec{F}^a_i+\vec{F}^H_i,
    \label{eq:monomer_motion}
 \end{equation}
 where 
 \begin{equation}
     U_i =U_{sp} + U_{bend} + U_{EV} \label{eq:potential}
 \end{equation}
   is the sum of the intramolecular potentials acting on the $i$th monomer. Likewise, $\vec{F}^a_i$ and $\vec{F}^H_i$ describe the active force and hydrodynamic force acting on the monomer. The latter force results in from the interaction of solvent molecules with the monomers and its effect will be incorporated by coupling the dynamics of monomers to theMCPD fluid particles as will be discussed in the subsequent section.
 
The bonding potential  between two consecutive monomers acting on the monomers of the chain is modelled by a harmonic potential of the form
\begin{equation}
        U_{sp}(\vec{R}_{i,i\pm1}) = \frac{k_s}{2}(|\vec{R}_{i,i\pm 1}| - \ell)^2,
        \label{eq:spring}
 \end{equation}
 in which the coefficient  $k_s$   describes the spring stiffness and $\ell$ is the equilibrium bond length. The bending potential is modeled as
 \begin{equation}
             U_{bend}(\Theta_{i}) = \kappa  (1-\cos{\Theta_{i}}),
          \label{eq:U_bend}
 \end{equation}
where  $\Theta_{i}$ denotes the angle between two consequent  bond vectors intersecting at bead $i$ defined as $\Theta_i= \cos^{-1} (\widehat{\vec{t}}_{i,i+1} \cdot \widehat{\vec{t}}_{i-1,i})$ with  $\widehat{\vec{t}}_{i,i+1}= \vec{R}_{i,i+1}/|\vec{R}_{i,i+1}|$ and $\vec{R}_{i,i+1}=\vec{R}_{i+1}-\vec{R}_{i}$. The coefficient  $\kappa$ is the bending stiffness determining the persistence length of the polymer and degree of flexibility of polymers.
Finally, the  excluded volume interactions are described by a truncated purely repulsive Lennard-Jones potential, {\it i.e},  the  Weeks-Chandler-Anderson (WCA) potential~\cite{WCA}.
 \begin{equation}
     U_{EV}(R_{ij}) = 
\begin{cases}
    4\epsilon\Bigg[\bigg(\frac{\sigma}{R_{ij}}\bigg)^{12}-\bigg(\frac{\sigma}{R_{ij}}\bigg)^6\Bigg] + \epsilon\text{,}\;  R_{ij} \leq 2^{1/6}\sigma \\
    0\text{,}\quad R_{ij} > 2^{1/6}\sigma,
\end{cases}
          \label{eq:polymodel}
 \end{equation}
 where $\sigma$ effectively sets the diameter of the monomers, and $R_{ij} =| \vec{R}_i - \vec{R}_j|$ is the distance between the $i$th and $j$th monomers.
 The active force $\vec{f}^a_i$ in Eq.~\eqref{eq:monomer_motion} is an external tangential force directed along the backbone of the polymer mimicking the effect of the tangential force exerted by the molecular motors on the polymer backbone. The  active force on each bead, except for the end monomers, is modeled as~\cite{Isele}:
\begin{equation}
   \vec{F}^a_i = \frac{f^a}{2 \ell}(\vec{R}_{i-1,i} + \vec{R}_{i,i+1})\\
     \label{eq:tangentactive}
\end{equation}
where $f^a$ sets the strength of the active force. The active force on the tail monomer is given by $\vec{F}^a_1=\frac{f^a}{2 \ell } \vec{R}_{1,2}$  and for the head monomer is $\vec{F}^a_{N_m} = \frac{f^a}{2 \ell}\vec{R}_{N_m-1,N_m}$. To incorporate  the hydrodynamic force on each monomer, we couple it to the MCPD particles as discussed in the following section.

 \subsection{Coupling of polymer dynamics to fluid}
 Applying an external tangential force to all the monomers creates a global fluid flow over time as the sum of the all forces is nonzero and is given by   
\begin{equation}
    \sum^{N_m}_{i=1}\bigg(-\nabla U_i+\vec{F}^a_i\bigg) = \sum^{N_m}_{i=1}\vec{F}_i^a=\vec{F}^a_{tot}(t),
        \label{eq:force_sum}
\end{equation}
where $\vec{F}^a_{tot}$ is the sum of all the tangential forces applied to the monomers.
The sum of forces induced by intramolecular interactions in the polymer cancel out due to Newton's third law, thus we can focus purely on the total active force $\vec{F}^a_{tot}$ itself.

\revise{For an externally-driven active polymer in real experimental situations with confining wall, the net active force $\vec{F}^a_{tot}$ exerted on the fluid is reflected by the confining container walls and induces a backflow.  The whole system has  zero momentum because the walls do not let the  external force to accelerate the center of mass of the whole system. In contrast, when we impose periodic boundary conditions in simulation, MCPD fluid particles can accelerate due to the external active force and 
result in a net fluid flow which accelerates the whole system in the absence of physical walls.  A globally moving system is an artifact of the periodic boundary conditions and not physical. To prevent a net fluid flow and artificial acceleration of center of mass of the whole system, we apply  a  decelerating backflow force to all the monomer and MPCD particles in such a way that the total momentum of the system (fluid plus active polymer) remains zero, {\it i.e.}, $\vec{F}^a_{tot}(t)+\vec{F}^b_{tot}(t)=0$.
We distribute backflow force equally among all MPCD particles and monomers proportional to their mass to produce a uniform acceleration opposite to the direction of total active force, similar to the case of  sedimenting soft colloids  exposed to a uniform gravitational force~\cite{Colloid_sedimentation} or externally driven active Brownian polymers~\cite{Martin-Gomez,Llahi}.      

More explicitly the  distribution of backflow forces acting on  the monomers $\vec{F}^b_{mon}$  and MCPD  particles $\vec{F}^b_{MPC}$ according to their masses~\cite{Martin-Gomez,Llahi} are given by}
\begin{equation}
    \vec{F}^b_{mon}(t) = -\frac{M}{mN + MN_m}\vec{F}^a_{tot}
    \label{eq:backflowmono}
\end{equation}
\begin{equation}
    \vec{F}^b_{MPC}(t) = -\frac{m}{mN + MN_m}\vec{F}^a_{tot}.
    \label{eq:backflowMPC}
\end{equation}
Here,  $M$ and $m$ are the masses of the monomers and MCPD  particles and $N_m$ and $N$ denote the number of monomers and MCPD  particles respectively. The magnitude of the backflow force is governed by the ratio between the particle mass and the total mass of the particles in the system. \\
Assuming that the timescale of changes of polymer conformation are larger than the collision time $\delta t$, the MCPD  fluid particle velocities $\vec{v}_k$ and positions $\vec{r}_k$  after a streaming step are given by
\begin{equation}
\begin{split}
    \vec{v}_k(t + \frac{1}{2}\delta t) = \vec{v}'_k(t) + \frac{\vec{F}^b_{MPC}}{2m} \delta t \\
    \vec{r}_k(t+\delta t) = \vec{r}_k + \delta t \vec{v}_k(t + \frac{1}{2}\delta t) \\
    \vec{v}_k(t + \delta t) = \vec{v}_k(t + \frac{1}{2}\delta t) + \frac{\vec{F}^b_{MPC}}{2m}\delta t.
\end{split}
\label{eq:Velocity-Verlet with force}
\end{equation}
  
   The velocities of  monomers in the presence of the fluid is determined by the solution of the following equation of motion:
\begin{equation}
    M\frac{\partial^2 \vec{R}_i}{\partial t^2} = M \frac{\partial \vec{V}_i}{\partial t }= -\nabla  U_i + \vec{F}_i^a + \vec{F}_{mon}^b,
    \label{eq:motion_updated}
\end{equation}
which incorporates coupling of monomers dynamics to the background flow. Here, $\vec{F}_{mon}^b$ is the backflow force applied to  each monomer given by Eq.~\eqref{eq:backflowmono}. Eq.~\ref{eq:motion_updated} is solved using the velocity Verlet integration scheme similar to Eq. \eqref{eq:Velocity-Verlet with force} but using the total force acting on each monomer instead of $F^b_{MPC}$. The SRD collision step now requires the fluid particles and monomers to undergo stochastic rotations relative to the centre of mass velocity of the collision cells. \revise{The stochastic collisions including both MPCD particles and the monomers  account for interactions of solute with solvent and  a build up of hydrodynamic interactions. }

The center of mass of which is given by
    \begin{equation}  \label{eq:comvelocity with objects2}
    \vec{v}_{cm} = \frac{\sum^{N_c}_{k=1}m\vec{v}_k + \sum^{N_c^m}_{j=1} M\vec{V}_j}{mN_c + MN_c^{m}},
   \end{equation}
\revise{where $N^m_c$ is the number of monomers in a collision cell}.  The stochastic rotation of MCPD  fluid particles and monomers take place  according to

\begin{align}
    \vec{v}'_k (t+\delta t)=&\vec{v}_{cm}(t+\delta t) \nonumber\\ +&\mathbf{\mathcal{R}}(\alpha)\cdot [\vec{v}_k(t+\delta t) - \vec{v}_{cm}(t+ \delta t)], \nonumber \\ \nonumber\\ 
\vec{V}'_i (t+\delta t)=&\vec{v}_{cm}(t+\delta t) +\nonumber\\ +&\mathbf{\mathcal{R}}(\alpha)\cdot [\vec{V}_i(t+\delta t) - \vec{v}_{cm}(t+ \delta t)], \nonumber
\end{align}
   where $\vec{v}_k(t+\delta t)$ is  determined  by Eq.~\eqref{eq:Velocity-Verlet with force} and $\vec{V}_i(t+\delta t)$  is given by the solution of Eq.~\eqref{eq:motion_updated} obtained by the velocity Verlet algorithm. This implies that a Stokeslet flow field by the active force appears for every monomer. 
   
   The discussed algorithm for external actuated active polymers with tangential active force is  implemented in Hoomd-blue \cite{hoomd1, hoomd2} with a home-made modification to include the active   and the backflow forces for the coupling of the MCPD  fluid  to the monomer dynamics.


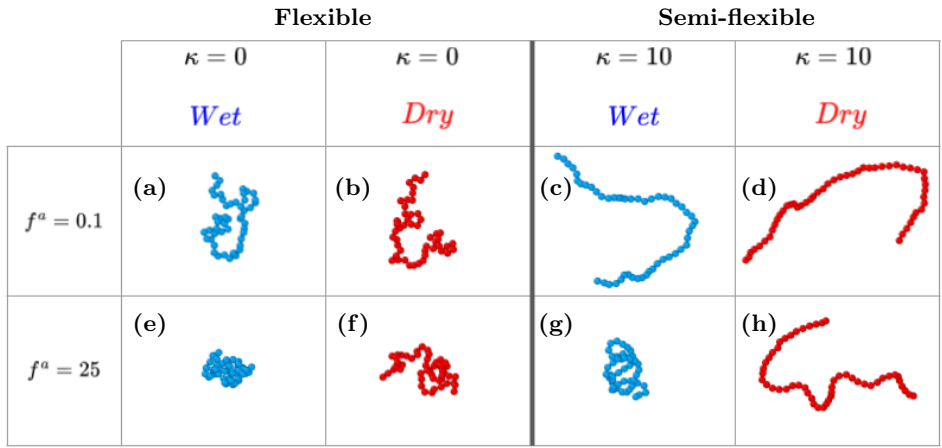
\begin{figure*}[ht]
    \centering
    \begin{tikzpicture}
            \draw (0,0) node[inner sep=0]{\includegraphics[width=0.7\linewidth ]{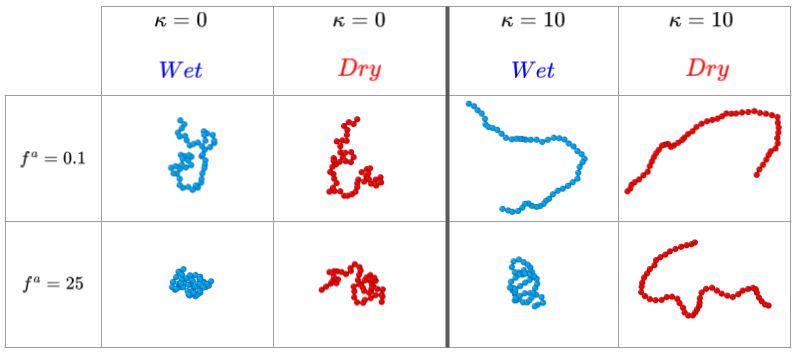}};
            \draw {(-2.,3.) node{\textbf{Flexible}}};
            \draw {(3.5,3.) node{\textbf{Semi-flexible}}};
            \draw {(-4.3,0.7) node{\textbf{(a)}}};
            \draw {(-1.6,0.7) node{\textbf{(b)}}};
            \draw {(1.1,0.7) node{\textbf{(c)}}};
            \draw {(3.8,0.7) node{\textbf{(d)}}};
            \draw {(-4.3,-1.1) node{\textbf{(e)}}};
            \draw {(-1.6,-1.1) node{\textbf{(f)}}};
            \draw {(1.1,-1.1) node{\textbf{(g)}}};
            \draw {(3.8,-1.1) node{\textbf{(h)}}};
    \end{tikzpicture}
    \caption{Snapshots of active polymers with $\kappa=0$ and $10$, $f^a=0.1$ and $25$ with and without HI.}
    \label{fig:Snaps}
\end{figure*}

\subsection{Dry active polymer simulations}
 To compare the effects of HI on the polymers, we also carried out Langevin dynamics simulations of dry  tangentially-driven polymers described by the  following equation of motion~\cite{Isele}. For each monomer $i$ we have:
 \begin{equation}
    M\frac{\partial^2 \vec{R}_i}{\partial t^2} = \gamma \vec{V}_{i} -\nabla  U_i + \vec{F}_i^a +  \vec{F}_i^r,
    \label{eq:motion_dry}
\end{equation}
  where $\gamma$ is the fluid friction coefficient per monomer. The potential energies $U_i$ are identical to those of wet active polymers as given by Eq.~\eqref{eq:potential}. The $\vec{F}_i^r$ is the random force on monomer $i$, chosen as a white noise with zero mean and has the correlation $\langle\vec{F}_i^r(t)\cdot\vec{F}_j^r(t')\rangle=6K_B T \gamma \delta_{ij} \delta(t-t')$. 
 
\subsection{Simulation parameters}
We choose the monomer diameter $\sigma$ as the unit of length, $k_B T$ as the unit of energy and  $\tau = \sqrt{m \sigma^2/ k_BT}$ as the unit of time. \revise{This leads to a dimensionless active force $f^a=F^a \sigma/ k_BT $, which can also be interpreted as monomeric Peclet number~\cite{Active_flexible}.} The  wet polymers   are simulated in a three-dimensional box with sides of length $L_{box} = 60$ and periodic boundary conditions.  The simulation box is split into collision cells of size $L_{cell} = 1\sigma$.  The mass of the MCPD  fluid particles is set to $m = 1$ and their collision time step  is set to $\delta t = 0.01 \sqrt{m \sigma^2/ k_BT}$.  The stochastic rotation angle is chosen as $\alpha = 130^{\circ}$. The number density of the MCPD  particles is $\langle N_c \rangle = 10$ leading to a total of  $N_s=2.16 \times 10^6$ MCPD  particles in the simulation box. 

\revise{The choice of $\delta t$, $N_c$ and $\alpha$ affect the viscosity of the resulting MPCD fluid $\eta = \eta_k + \eta_c$, which includes kinetic  $\eta_k$ and collisional  $\eta_c$ contributions~\cite{MPCD_viscosity,winkler2009stress}.
  The kinetic viscosity $\eta_{k}$ is directly proportional to $\delta t$~\cite{MPCD_viscosity} and is dominant at large timesteps   representing gas-like behaviour  of a fluid, whereas the collision viscosity $\eta_{c}$ is inversely proportional to $\delta t$. The latter predominates the total viscosity for  small time steps mimicking viscous fluid behaviour~\cite{Schmidt,Thermostat-viscosity}. It is given by~\cite{MPCD_viscosity}
    \begin{equation}
       \eta_c = \frac{\langle N_c\rangle \; m }{18 \; L_{cell} \;\delta t} \left( 1 - \cos(\alpha) \right) \left( 1 - \frac{1}{\langle N_c\rangle} \right)
   \end{equation}
Our  choice of MPCD parameters above yields $\eta_c \gg \eta_k$, hence the viscosity is $\eta \approx \eta_c \approx 82.1 \frac{\sqrt{m k_B T}}{\sigma^2}$.}


We investigate conformation and dynamics of polymers with  $N_m = 50$ monomer of two different values of bending rigidity,  $\kappa= 0$  and $10$, leading to fully flexible  and semi-flexible  polymers. The excluded volume interaction strength is set to  \revise{$\epsilon =1$}.
We choose the monomer mass to be $M = m\langle N_c \rangle = 10$.    The magnitude of the dimensionless active force varies in the range  $0.1<f^a< 25$. We set  bond  rest length $\ell$ to unity and the spring constant of the bonds is chosen as $k_s = (10+2 f^a) \times 10^3$  to increase linearly with the activity level to prevent bond overstretching and to ensure the bond lengths are always close to unity with an error margin of $\%5$, \revise{See Appendix A, for the distribution of bond lengths of flexible and semiflexible active polymers}. The monomer  positions and velocities are updated with time steps of $\Delta t = 0.001=\delta t/10$.

 We simulate 150 individual wet polymers. For each polymer, we first generate a three dimensional random walk of length $N_m$ and step size $\ell$. We use this sequence as the initial positions of the monomers. Once the polymer is created, the system is simulated for $10^7$-$10^8$ time steps to  reach a steady state depending on the value of $f^a$. Then we run a production simulation of $2\times 10^7$ - $2\times 10^8$ time steps, depending on the activity level.
 \revise{To ensure that our simulation results are not affected by the chosen box size and potential hydrodynamic self-interactions, we also ran additional simulations with bigger box size $L_{box}=100$ for $f^a=0.1$ and 25 for the semiflexible case, which led to identical static and dynamical features as discussed in more details in Appendix.\ref{APP:L100}.}

    Simulations of dry active polymers, conducted for comparison with their wet counterparts, involve 120 individual dry active polymers described by Eq.~\eqref{eq:motion_dry} in a box with size $L_{box}=200$, while all other parameters being identical to those of wet active polymer model. 
    The friction coefficient of monomers in dry active polymer simulation is chosen to be $\gamma=3 \pi \eta \sigma \approx 773 $.
This leads to a ratio of $M/\gamma \approx 0.013$ which ensures that the dynamics of monomers are overdamped and inertial effects~\cite{Fazelzadeh} remain negligible.  To reach the steady state, we use $10^7$ to $10^9$ time steps, depending on the activity level. For the production run, twice the number of time steps used to reach the steady state are used. The time step for dry simulations is $\Delta t' = 0.002$.



 \section{Conformational features of wet active polymers \label{Sec:conformation}}
 We first visually inspect the effects of  HI on the conformation of the chains.   Fig.~\ref{fig:Snaps} shows typical representative snapshots of active polymers with and without  HI for two bending stiffness values $\kappa=0$ and 10 at low and high active forces given by $f^a=0.1$ and 25 (see videos S1-S8).  With the increase of activity level, both wet and dry active polymers exhibit more shrunken conformations at the higher active force of $f^a = 25$, with the wet active polymers appearing slightly more shrunken.   Moving to the semiflexible chains with $\kappa=10$,  the overall conformation of the dry and the wet  chains at the low activity level of $f^a=0.1$ look alike. In contrast, at high active force of $f^a=25$,  fluid-mediated interactions induce a notable shrinkage in wet active polymers compared to their dry counterparts.

 \begin{figure}[t]
    \centering
    \begin{tikzpicture}
            \draw (0,0) node[inner sep=0]{\includegraphics[width=0.9\linewidth ]{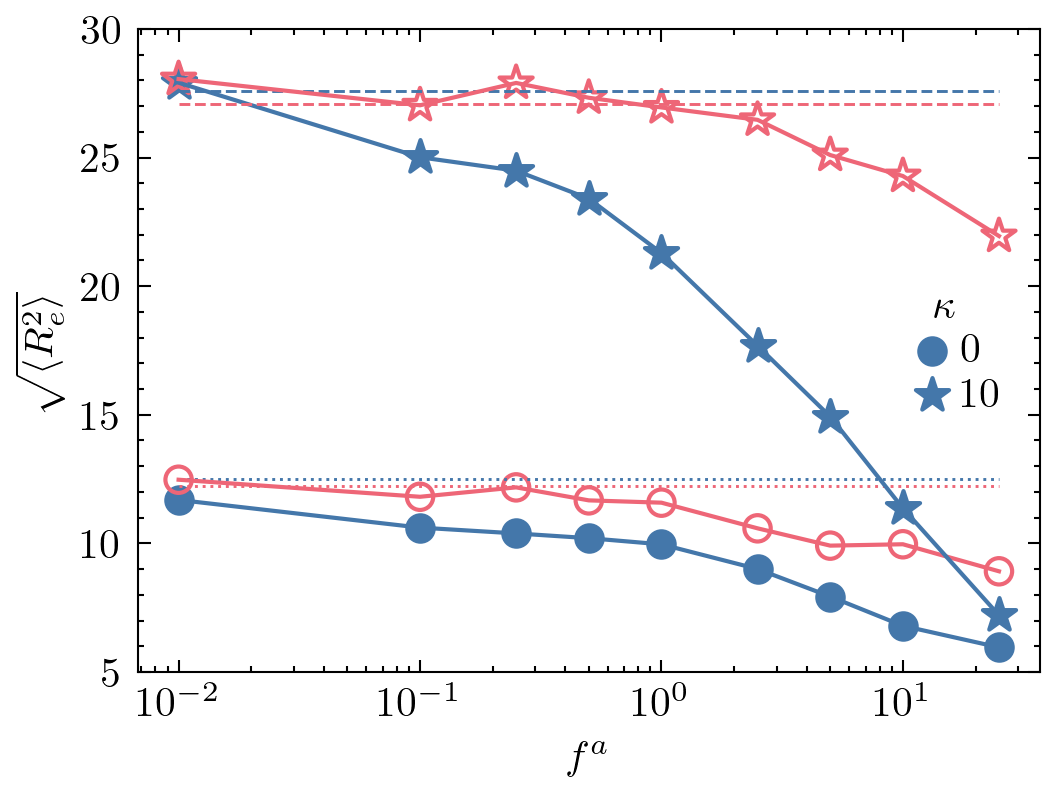}};
    \end{tikzpicture}
    \caption{\revise{The root of the mean squared end-to-end distance $\sqrt{\langle R_e^2 \rangle}$ versus active force $f^a$ for chains with $\kappa=0$ and $10$. Closed and open symbols represent data for chains with and without hydrodynamic interaction, respectively. The dotted and dashed lines show the values at $f^a=0$ for $\kappa=0$ and $10$, respectively.}}
    \label{fig:Re}
\end{figure}
\begin{figure}[t]
    \centering
    \begin{tikzpicture}
            \draw (0,0) node[inner sep=0]{\includegraphics[width=0.9\linewidth ]{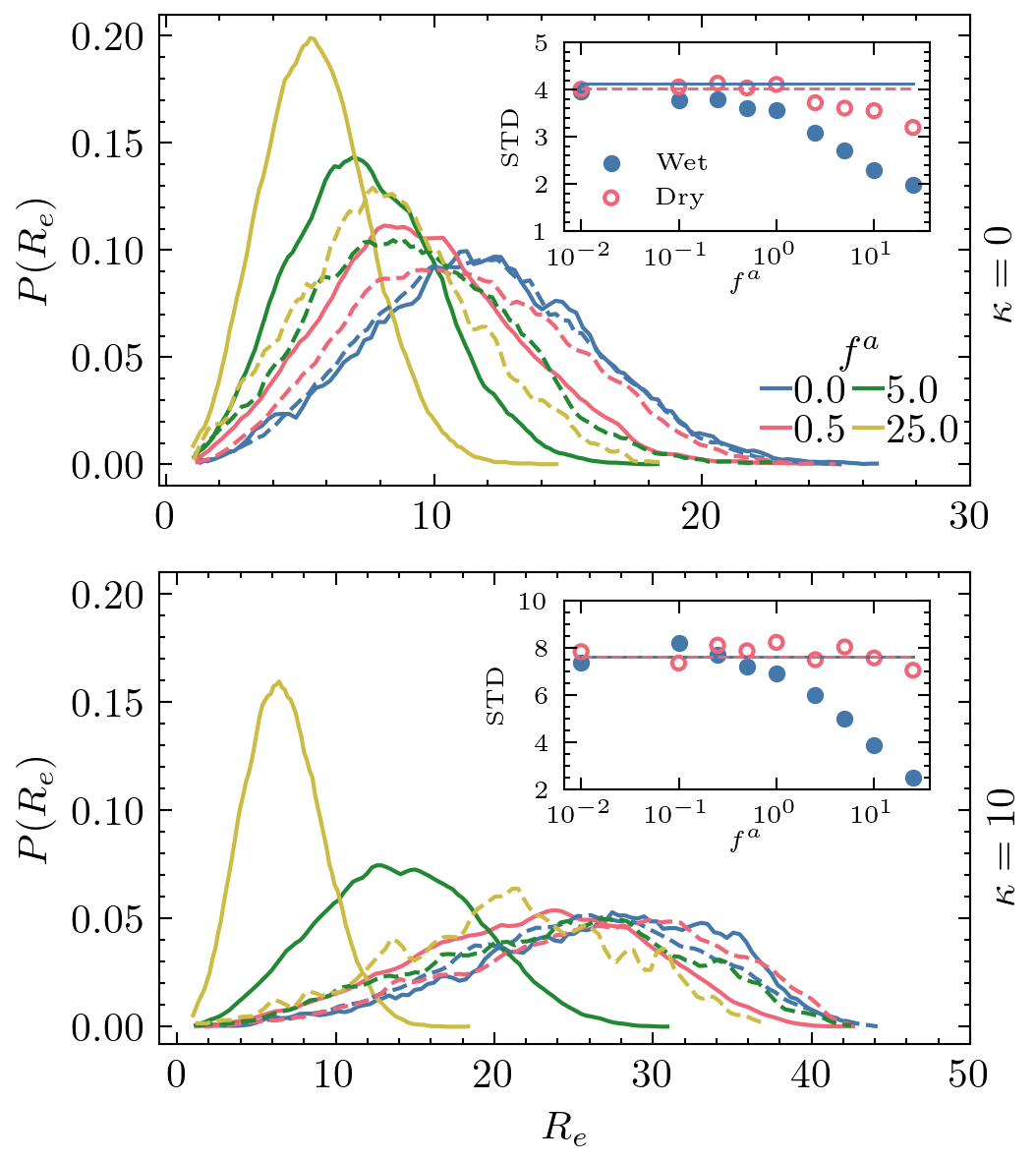}};
            \draw {(-2.4,3.75) node{\textbf{(a)}}};
            \draw {(-2.4,-0.4) node{\textbf{(b)}}};
    \end{tikzpicture}
    \caption{\revise{The probability distribution function of the end-to-end distance $P(R_e)$ for different active forces $f^a$ for chains with $\kappa=0$ and $10$. Solid and dashed lines represent data for chains with and without hydrodynamic interaction, respectively. The insets show the standard deviations (STD) with respect to activity. The horizontal solid and dashed lines in the insets show the values at $f^a=0$ for wet and dry polymers respectively. } }
    \label{fig:PDFRe}
\end{figure}

 \subsection{End-to-end distance}
 To have a quantitative assessment  of hydrodynamic effects on the mean size of active polymers, we examine the root of the mean squared end-to-end distances of the chains, \emph{i.e.}, $\sqrt{\langle R_e^2\rangle} = \sqrt{\langle |\vec{R}_{N_m} - \vec{R}_1|^2\rangle}$. Fig.~\ref{fig:Re} depicts the $\sqrt{\langle R_e^2\rangle}$ against activity for dry and wet chains with $\kappa=0$ and $10$, \revise{where the dotted and dashed lines show the corresponding $\sqrt{\langle R_e^2\rangle}$ of passive polymers.  First, we note that  for the  passive and very low activity limit $f^a=0.01$, the $\sqrt{\langle R_e^2\rangle}$ of wet and dry polymer are equal within the error bars, confirming that for HI interactions do not affect static features of passive polymers. Upon increasing active force, for} flexible chains with $\kappa = 0$, we observe slightly smaller mean end-to-end distance values for wet polymers compared to their dry counterparts. As the activity level increases, the difference between the two becomes more pronounced. Looking into the $\sqrt{\langle R_e^2\rangle}$ of  semiflexible chains with $\kappa=10$ in Fig.~\ref{fig:Re}, we observe a  more notable shrinkage of chains  compared to their flexible counterparts, specially at larger active forces $f^a>1$ where the 
 mean size of wet active polymers is about twice smaller than the dry ones.

 To better assess the impact of HI on the conformation of active polymers, we examine the probability distribution function (PDF) of $R_e$, presented  in Fig.~\ref{fig:PDFRe} with \revise{the insets showing the corresponding standard deviation (STD) of each PDF}.  For both  flexible and semiflexible polymers, the distribution of the end-to-end distance of wet active polymers shifts towards smaller $R_e$ values relative to their dry counterparts and \revise{the $P(R_e)$ of wet active chains become narrower as their STD drops with increasing activity.  This effect is particularly noticeable in the probability distribution $P(R_e)$ of wet active polymers at $f^a=25$.} In both flexible and semiflexible chains, narrow peaks emerge at lower $R_e$ values for wet active polymers, contrasting sharply with their dry counterparts. In the case of semiflexible chains, the HI-induced shrinkage of conformation is substantial enough to alter the overall shape of wet active polymers from a polymer-like conformation similar to that of dry polymers into a helix-like conformation, as demonstrated in Fig.~\ref{fig:Snaps}.

 \subsection{Interachain correlations}
 Having investigated the effects of HI on the statistical properties of end-to-end distance  of active polymers, we study its impact on \revise{intrachain correlations by examining the bond-bond correlation function.}

\begin{figure}[t]
    \centering
    \begin{tikzpicture}
            \draw (0,0) node[inner sep=0]{\includegraphics[width=1\linewidth ]{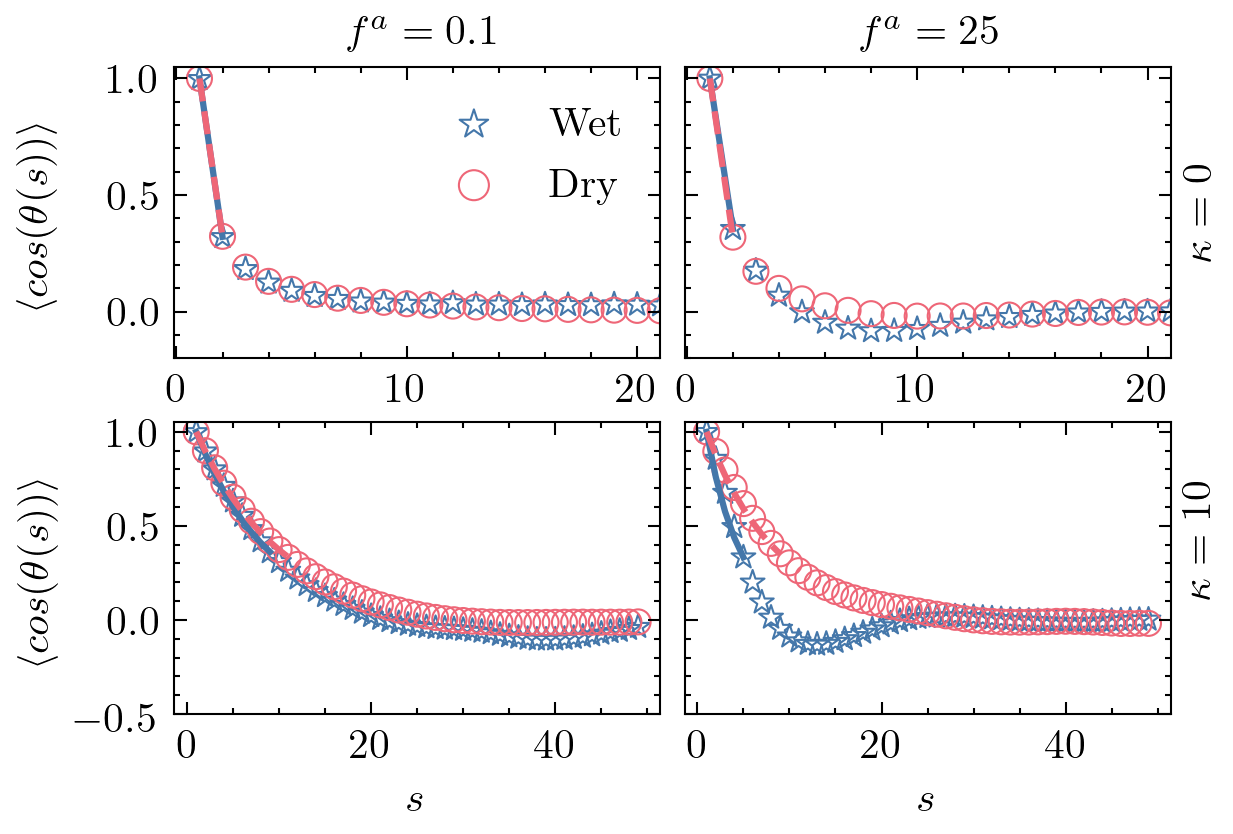}};
            \draw {(-0.1,1.0) node{\textbf{(a)}}};
            \draw {(3.5,1.0) node{\textbf{(b)}}};
            \draw {(-0.1,-0.9) node{\textbf{(c)}}};
            \draw {(3.5,-0.9) node{\textbf{(d)}}};
    \end{tikzpicture}
    \caption{The  bond-bond correlation function $\langle \cos(\theta(s)) \rangle$ versus curvilinear distance $s$ for wet and dry polymers for the bending stiffness  $\kappa=0$ and active forces of a) $f^a=0.1$ and b) $f^a=25$ and the bending stiffness of $\kappa=10$ with active forces of c) $f^a=0.1$ and d) $f^a=25$. The solid lines show the fit to an exponential function for $s<s_e$.}
    \label{fig:bbcorr}
\end{figure}
\begin{figure}[t]
    \centering
    \begin{tikzpicture}
            \draw (0,0) node[inner sep=0]{\includegraphics[width=0.9\linewidth ]{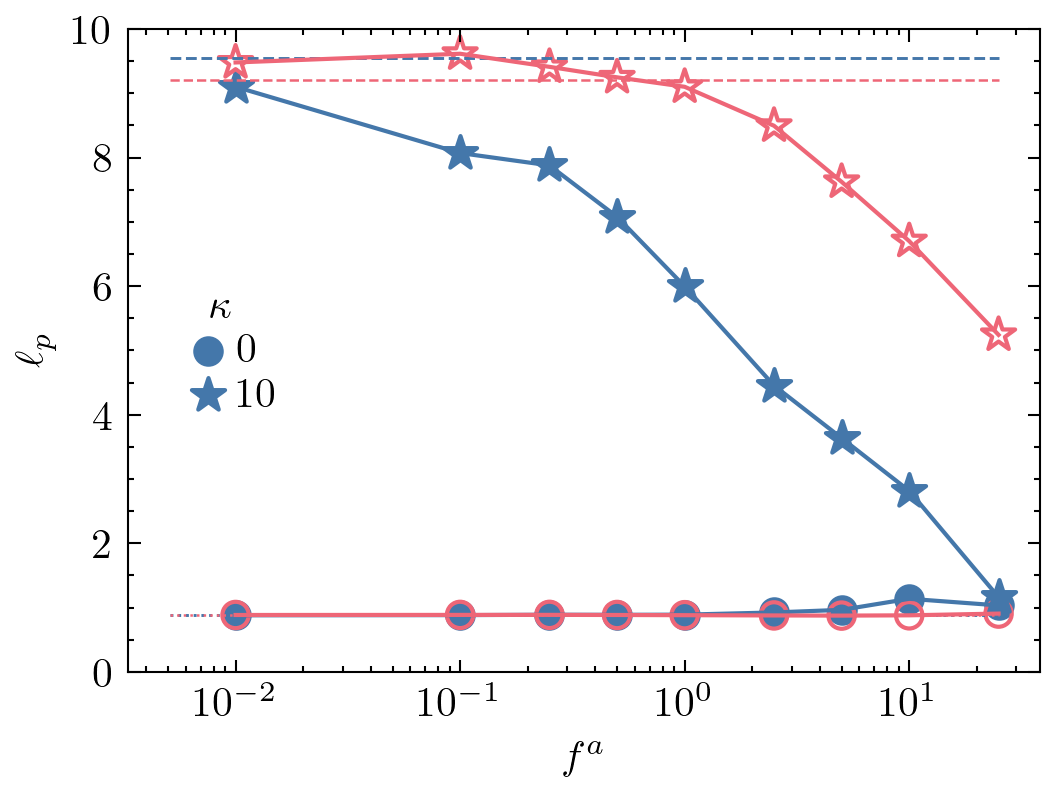}};
    \end{tikzpicture}
   \caption{ \revise{The effective persistence length  $\ell_p$ versus active force $f^a$ for chains with $\kappa=0$ and $10$. Close and open symbols represent data for chains with and without hydrodynamic interaction, respectively. The dotted and dashed lines show the values at $f^a=0$ for $\kappa=0$ and $10$, respectively.}}
    \label{fig:Lp}
\end{figure}

The bond-bond correlation function is defined as $\langle cos(\theta(s)) \rangle$, where $\theta(s)$ is the angle between two bond vectors with a curvilinear distance $s$. Fig.~\ref{fig:bbcorr} shows the bond-bond correlation functions of wet and dry chains with low and high activities at two different degrees of flexibility. For flexible polymers with $\kappa=0$, the correlations drop to zero at small curvilinear distances ($s<5$), regardless of the activity of the chains or inclusion of HI, see Fig.~\ref{fig:bbcorr}(a) and (b). However, at the high active force of $f^a=25$, $\langle cos(\theta(s)) \rangle$ of wet active chains exhibits a slightly negative dip reflecting anti-correlation of bond vectors at curvilinear distances around $s\approx 10 $ and chain folding.
For semiflexible chains with $\kappa=10$, at the lower activity level with $f^a=0.1$, the dry and wet polymers exhibit very similar bond-bond correlation functions with  the $\langle cos(\theta(s)) \rangle$ decaying in slightly shorter distances than their dry counterparts. The impact of HI is most remarkable at the high activity of $f^a=25$, shown in Fig.~\ref{fig:bbcorr}(d), where the bond-bond correlation of the wet  and dry  active chains are notably different. While the bond correlation of the dry active chain decays monotonically to zero, the bond-bond correlation functions of wet active polymers exhibits a negative cusp at short curvilinear distances ($s \approx 10$ comparable to their bare persistence length $\ell_p^0=10$) reflecting the folding of active polymers at this scale resulting in helix-like coiling of chains as can be seen from Fig.~\ref{fig:Snaps} (g). 

To quantify the combined effects of activity and HI on  effective stiffness of the polymer backbone, we obtain an effective persistence length $\ell_p$ defined with the following protocol. First, we define $s_e$ as the shortest curvilinear distance at which the correlation becomes  smaller than $e^{-1}$, \emph{i.e},  $\langle cos(\theta(s_e))\rangle \leq e^{-1}$. We then fit the bond correlation data in the range $0 \leq s \leq s_e$ by an exponential function $\langle cos(\theta(s))\rangle=\exp(-s/\ell_p)$ as shown in Fig.~\ref{fig:bbcorr} and  evaluate the persistence length. Fig.~\ref{fig:Lp} shows the effective persistence length against activity for dry and wet chains with $\kappa=0$ and $10$. We recall that according to the worm-like chain model~\cite{Binder_2020}, the bare persistence length of passive dry chains is $\ell_p^0=\kappa \sigma / (k_BT)$.\\
As can be seen from Fig.~\ref{fig:Lp}, \revise{for fully flexible chains with $\kappa=0$, bond-bond correlations decay very rapidly as expected and the effective persistence lengths  of wet and dry active chains are of the order of unity independent of  their activity levels. } For semiflexible chains with $\kappa=10$, \revise{the effective persistence lengths $\ell_p$ of the wet polymers become shorter than those of their dry counterparts as the activity increases. For both wet and dry chains, the $\ell_p$ values at $f^a=0$ and at the lowest activity ($f^a=0.01$) are close to the bare persistence length $\ell_p^0=10$, see the dashed lines in Fig.~\ref{fig:Lp} for the $\ell_p$ of passive chains.} The effective persistence lengths decrease with activity for both dry and wet chains, because high active forces  increase the likelihood of the polymer backbone overcoming bending energy barriers to adopt more flexible conformations. Consistent with the results of the end-to-end distance, the $\ell_p$ of wet semiflexible chains exhibits a significant decrease at high activity levels. For instance, at a high active force of $f^a=25$, the effective persistence length of wet semiflexible chains matches that of flexible wet active polymers with $\kappa=0$. This finding, along with the end-to-end distance results, underscore the significant impact of HI on the conformation of highly active semiflexible chains.

\begin{figure*}[ht]
     \centering
    \begin{tikzpicture}
            \draw (-4.5,0) node[inner sep=0]{\includegraphics[width=0.495\linewidth ]{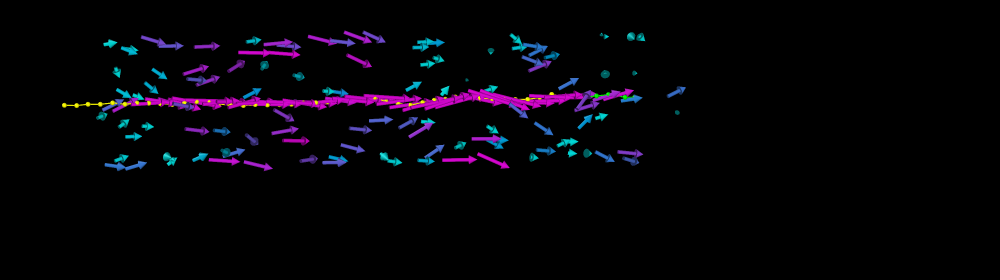}};
            \draw {(-4.5,-2.2) node{\includegraphics[width=0.495\linewidth ]{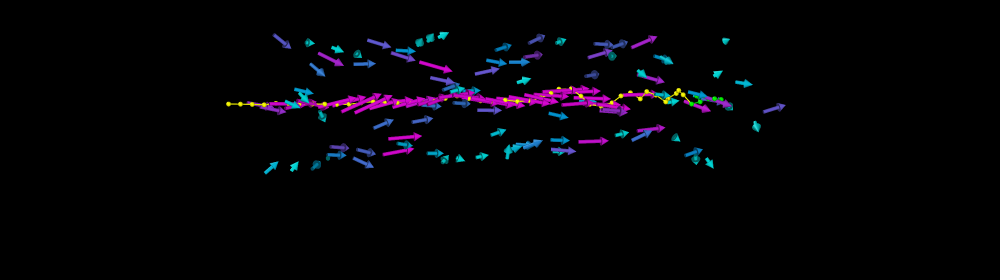}}};
            \draw {(-4.5,-4.4) node{\includegraphics[width=0.495\linewidth ]{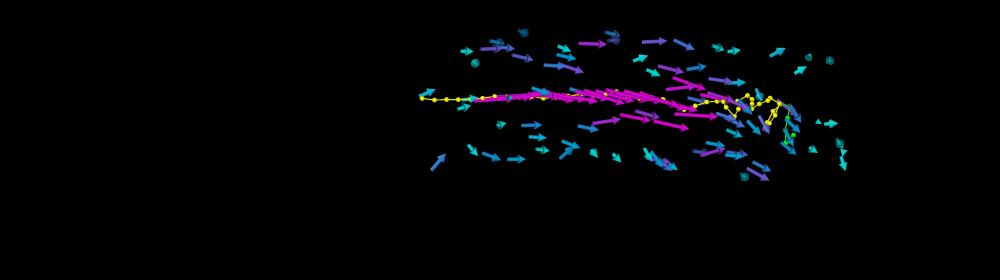}}};
            
            \draw {(4.5,0) node{\includegraphics[width=0.495\linewidth ]{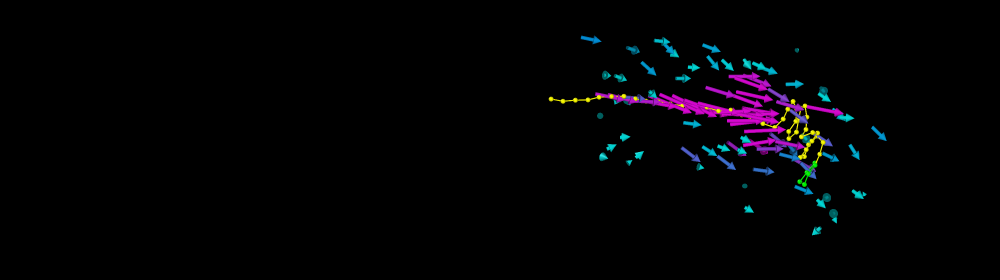}}};
            \draw {(4.5,-2.2) node{\includegraphics[width=0.495\linewidth ]{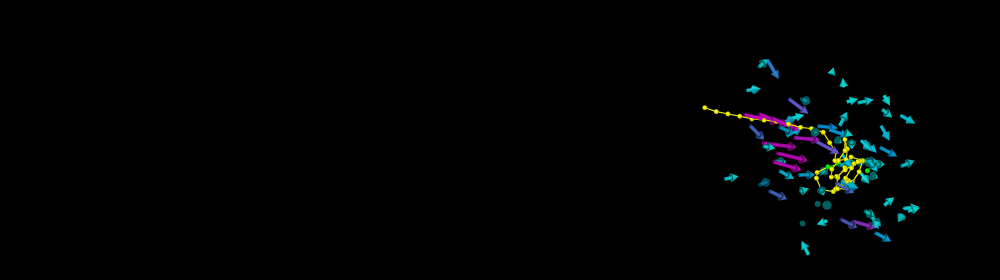}}};
            \draw {(4.5,-4.4) node{\includegraphics[width=0.495\linewidth ]{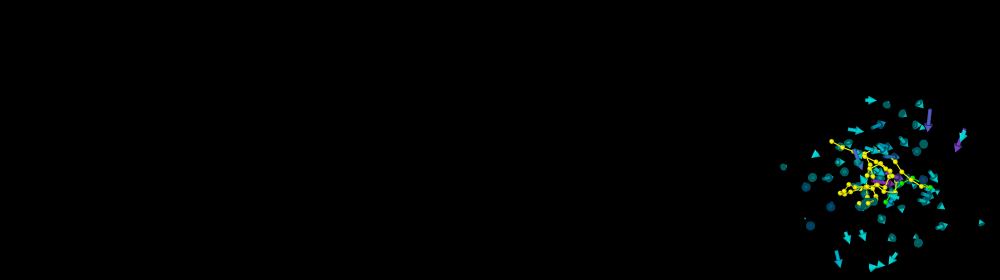}}};
            \draw {(-8.,0.8) node{{\color{white}\textbf{(a) $t=0$}}}};
            \draw {(-8.,-1.5) node{{\color{white}\textbf{(b) $t=30$}}}};
            \draw {(-8.,-3.6) node{{\color{white}\textbf{(c) $t=60$}}}};
            \draw {(+1.1,0.8) node{{\color{white}\textbf{(d) $t=90$}}}};
            \draw {(1.1,-1.5) node{{\color{white}\textbf{(e) $t=120$}}}};
            \draw {(1.1,-3.6) node{{\color{white}\textbf{(f) $t=150$}}}};
            \draw {(0,-6.) node{\includegraphics[width=0.2\linewidth ]{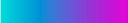}}};
            
            \draw {(-2.15,-6.) node{$0.02$}};
            \draw {(2.2,-6.) node{$0.2$}};
            \draw {(0,-6.6) node{\textbf{Average fluid \revise{speed}}}};

    \end{tikzpicture}
    \caption{Snapshots of wet active polymer with $\kappa=10$ and activity of $f^a=25$ at different times, showing the average flow field at monomer positions and their vicinity. The green segment shows the head of the polymer and the rest of the monomers are shown in yellow. The color of the arrows encodes the \revise{speed} of the flow field. }
    \label{fig:Mechanism}
\end{figure*}
 
 \subsection{\revise{Enhancement of shrinkage with HI}}
As we have already seen from Figs.~\ref{fig:Snaps}(g) and \ref{fig:Re},  \revise{HI lead to greater shrinkage of semiflexible polymers at high activity levels compared to that observed in their dry counterparts. This feature is a manifestation of non-linear coupling of bending elasticity, activity and hydrodynamic effects for non-equilibrium polymers.} To understand the underlying mechanism of \revise{the enhanced} shrinkage, we investigate the flow field in the vicinity of the backbone of an active chain with $\kappa=10$ and $f^a=25$,  initialized as an straight line, \revise{ where the springs have their equilibrium rest length}.\\
Fig.~\ref{fig:Mechanism} shows snapshots of this chain and the \revise{smoothed coarse-grained} flow field at the monomer positions and their vicinity in a time series (see video S9). \revise{The instantaneous  smoothed flow field of a point of interest $P=(x,y,z)$ at any time is measured by averaging over the velocity of all fluid particles within a sphere of radius $3 L_{cell}$ centered at $P$}. At $t=0$ when the  chain is in a rod-like conformation, \revise{since the activity is high we observe a strong flow} field nearly  parallel to the backbone but heterogeneous in strength  \revise{ with the given spatial distribution of active forces, see Eq.~\eqref{eq:tangentactive}. The initial flow field for this conformation can be evaluated by summing the flow fields of stokeslets of equal strengths, apart from the end head and tail segments, which are all parallel to the chain direction, see Appendix. C. 
Doing so   we find that the flow field is largest in the middle of the straight chain where  active forces are distributed on both sides of the monomers. In contrast, at the head and tail segments, where forces are distributed only on one side of the monomer, the flow field is weakest.

This heterogeneous distribution of flow field along the polymer backbone renders the straight conformation  mechanically unstable. Consequently, the head segment moves slower than the rest of the polymer, while the tail does not experience the same slowdown, as it is pulled forward by the faster-moving segments ahead of it. 
This hydrodynamics instability coupled to dynamical active force rule, bending elasticity and thermal fluctuations  lead to evolution of the  conformation as presented in Fig.~\ref{fig:Mechanism}(b)-(f).}  Because the rear segments move faster than the front segments, they  create a compressional flow which pushes the rear \revise{segments} towards the front ones resulting in chain buckling, as illustrated in Fig.~\ref{fig:Mechanism}(c) and (d).
  Finally, the difference in fluid velocity between the buckled and straight regions of the polymer further compresses its conformation until the entire structure becomes \revise{coiled}, as seen in Fig.~\ref{fig:Mechanism}(e) and (f). The general pattern in the motion of semiflexible wet active polymers consists of \revise{activity-induced flow heterogeneity generates compressional flows, causing the active polymer to buckle, leading to local folding and coiling. This is followed by flow fluctuations at the head segment, allowing the head to escape the buckled conformation, resulting in the gradual unwinding of the coiled structure,}  see video S8.
 



 \section{Dynamical features of wet active polymers \label{Sec:dynamics}}
Having discussed the conformational properties of wet active polymers, next we investigate the effects of HI on their dynamical features. 
 
   \begin{figure}[t]
    \centering
    \begin{tikzpicture}
            \draw (0,0) node[inner sep=0]{\includegraphics[width=1\linewidth ]{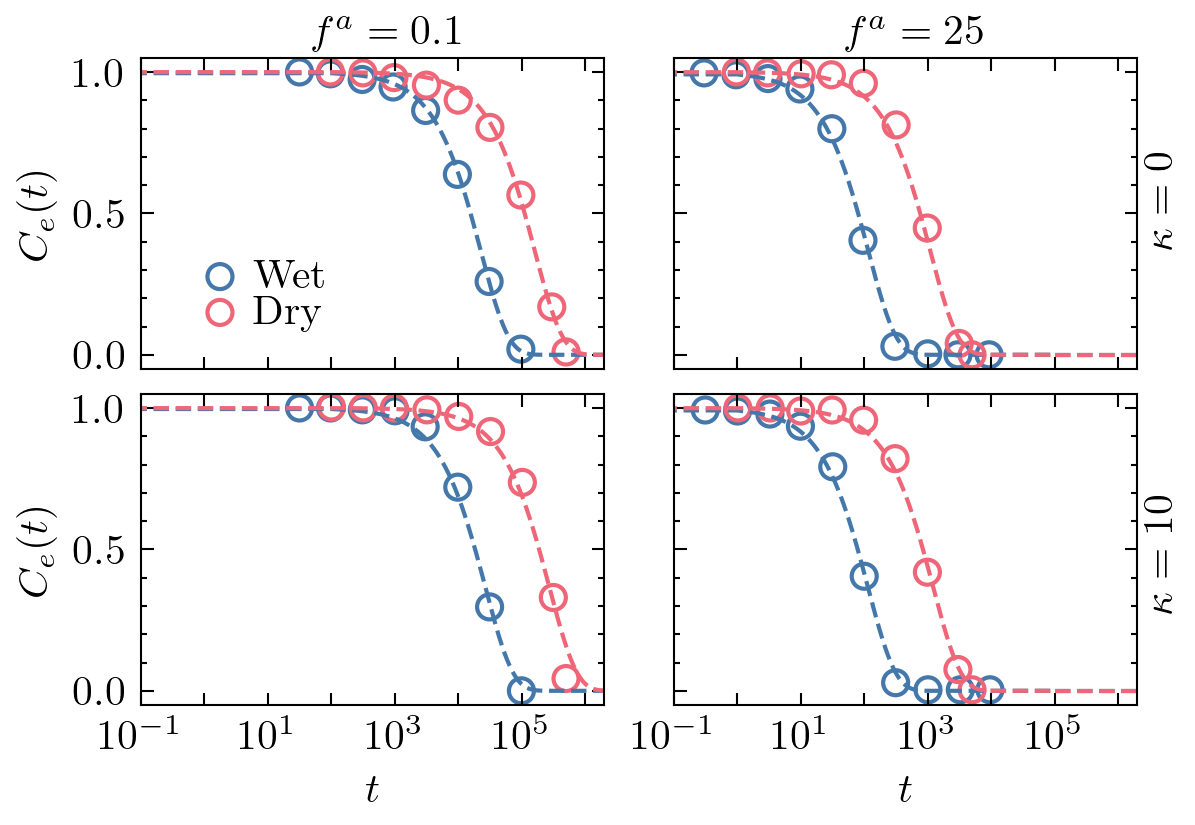}};
            \draw {(-3.,2.) node{\textbf{(a)}}};
            \draw {(.9,2.) node{\textbf{(b)}}};
            \draw {(-3,-0.5) node{\textbf{(c)}}};
            \draw {(.9,-0.5) node{\textbf{(d)}}};
    \end{tikzpicture}
    \caption{\revise{The normalized end-to-end vector auto-correlation $C_e(t)$ as a function of lag time $t$ for wet and dry polymers with bending stiffness of $\kappa=0$ and active forces of a) $f^a=0.1$ and b) $f^a=25$ and bending stiffness of $\kappa=10$ with active forces of c) $f^a=0.1$ and d) $f^a=25$. The lines depict exponential fits to the data for the time interval where $1\leq C_e(t) \leq e^{-1}$.}}
    \label{fig:ct}
\end{figure}

\begin{figure}[t]
    \centering
    \begin{tikzpicture}
            \draw (0,0) node[inner sep=0]{\includegraphics[width=0.9\linewidth ]{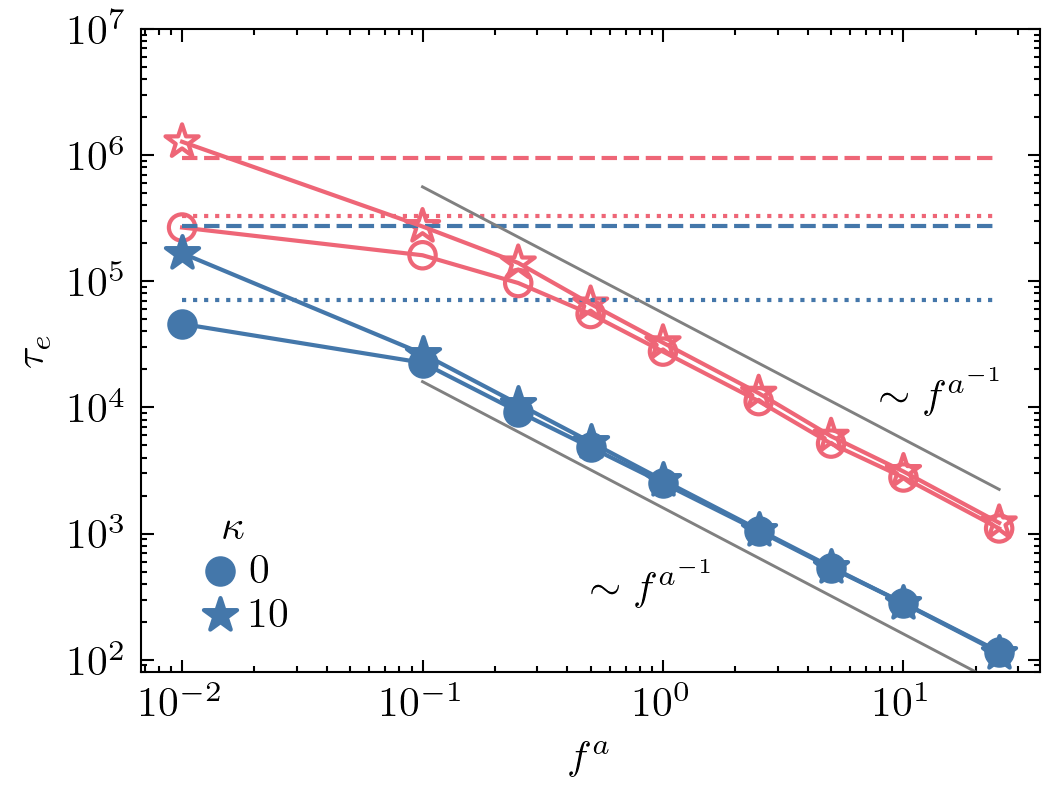}};
    \end{tikzpicture}
    \caption{\revise{The end-to-end vector relaxation time  $\tau_e$ versus active force $f^a$ for chains with $\kappa=0$ and $10$. Close and open symbols represent data for chains with and without hydrodynamic interaction, respectively. The dotted and dashed lines show the values at $f^a=0$ for $\kappa=0$ and $10$, respectively.}}
    \label{fig:taue}
\end{figure}

 \begin{figure*}[t]
     \centering
    \begin{tikzpicture}
            \draw (0,0) node[inner sep=0]{\includegraphics[width=0.8\linewidth ]{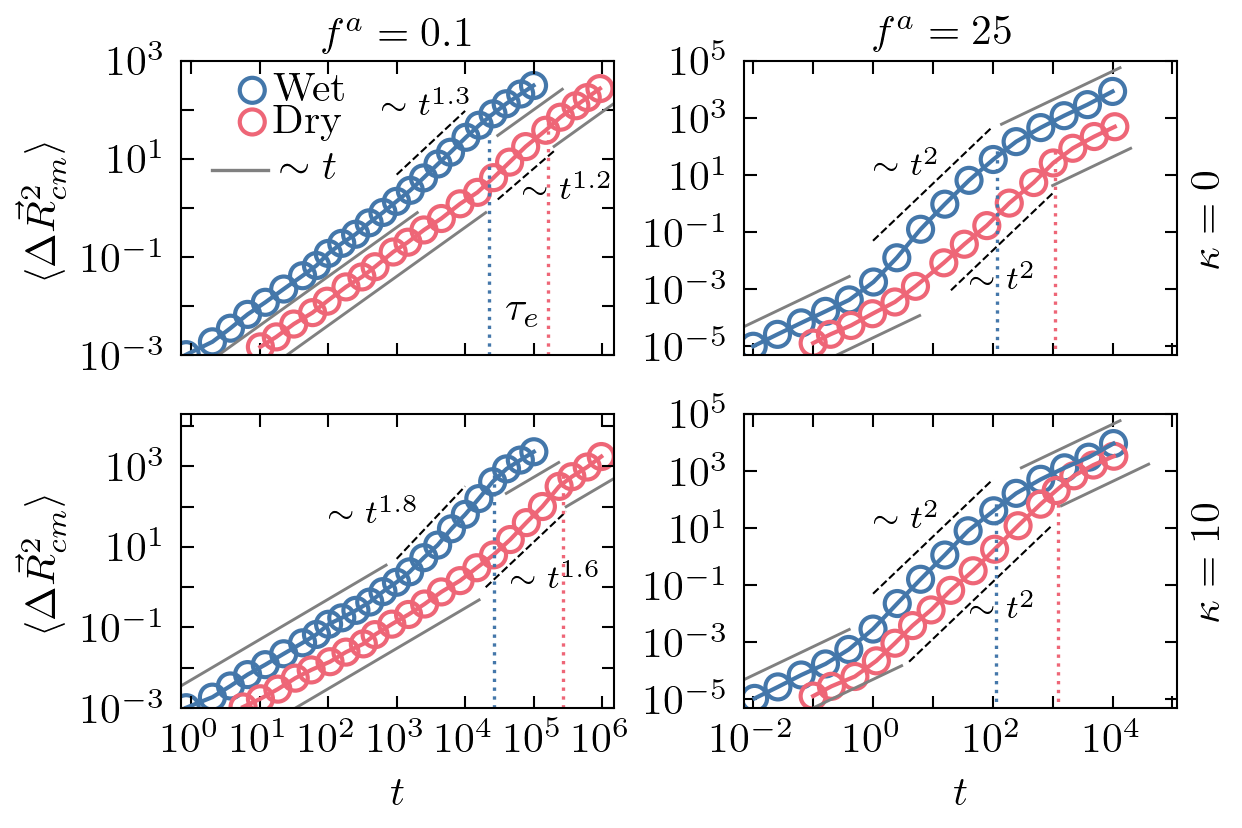}};
            \draw {(-4.6,2.2) node{\Large{\textbf{(a)}}}};
            \draw {(2.2,2.2) node{\Large{\textbf{(b)}}}};
            \draw {(-4.6,-2.) node{\Large{\textbf{(c)}}}};
            \draw {(2.2,-2.) node{\Large{\textbf{(d)}}}};
    \end{tikzpicture}
    \caption{The mean squared displacement of the center of mass $\langle \Delta \vec{R}_{cm}^{2}\rangle$ as a function of lag time $t$ for wet and dry polymers with bending stiffness of $\kappa=0$ and active forces of a) $f^a=0.1$ and b) $f^a=25$ and bending stiffness of $\kappa=10$ with active forces of c) $f^a=0.1$ and d) $f^a=25$. The gray solid lines show diffusive regimes ($\langle \Delta \vec{R}_{cm}^{2}\rangle \sim t$), while the black dashed lines show super diffusive regimes ($\langle \Delta \vec{R}_{cm}^{2}\rangle \sim t^{\alpha}$) with their corresponding exponent $\alpha>1$. The vertical dotted lines show the corresponding relaxation times $\tau_e$ of wet (blue) and dry (red) active polymers.}
    \label{fig:msd}
\end{figure*}

\subsection{Orientational dynamics}
We start by examining the effects of HI on orientational dynamics of active polymers. Since a tangentially-driven polymer is polar (head-tail asymmetry), we define the end-to-end vector as $\vec{R}_e(t)=\vec{R}_{N_{m}}(t)-\vec{R}_{1}(t)$ where $\vec{R}_{N_m}$ corresponds to the head of polar active polymer. In tangentially-driven active polymers, the net active self-propulsion force on the center of mass is proportional to the end-to-end vector, $\vec{F}^{a}(t)=\sum_{i=1}^{N}\vec{f}_{i}^a(t)=f^a \Vec{R}_{e}(t)/\ell$. Thus, we characterize the orientational dynamics of the polymers  by the normalized time auto-correlation function (TACF) of their end-to-end vector,
\begin{equation}
    C_{e}(t)= \dfrac{\langle \Vec{R}_{e}(0).\Vec{R}_{e}(t) \rangle}{\langle \Vec{R}_{e}^2 \rangle}.
    \label{eq:Ce}
\end{equation}

In Fig.~\ref{fig:ct} we present the normalized TACF of the end-to-end vector of wet and dry polymers with flexible and semiflexible backbones at low and high activity levels. Generally, the plots demonstrate that the HI lead  to a faster decay of the TACF of end-to-end vectors of wet polymers in comparison with their dry counterparts. Furthermore, by increasing the activity, we observe faster decays in the orientational dynamics of polymers, compare Figs.~\ref{fig:ct} (a) and (b) for flexible polymers and Figs.~\ref{fig:ct} (c) and (d) for semiflexible polymers.  While we do observe a notable impact of HI on the decay of $C_e$, the level of flexibility appears to have minimal influence on the decay of the orientational TACF. \\

To quantify the influence of  HI on the decay times of orientational dynamics, we introduce a characteristic orientational relaxation time, $\tau_e$, by the following procedure. First, we define $t_e$ as the shortest lag time at which the normalized TACF becomes equal to $e^{-1}$. Subsequently, we fit the data within the range $0 \leq t \leq t_e$ using an exponential function $C_{e}(t)=\text{exp}(-t/\tau_e)$, shown by dashed lines in Fig.~\ref{fig:ct}, to determine the orientational relaxation time, $\tau_e$. Figure~\ref{fig:taue} presents the extracted orientational relaxation times versus active force for both wet and dry polymers with flexible and semiflexible backbones. Consistent with our previous observations, we find no significant difference in $\tau_e$ among chains with varying flexibility levels. However, wet active polymers exhibit notably shorter relaxation times, nearly one order of magnitude smaller, indicating the rapid decay of their end-to-end vector time autocorrelation function (TACF). Additionally, the relaxation times of wet chains follow a $1/f^a$ scaling behavior, akin to the findings of previous studies of dry tangentially-driven active polymers~\cite{Active_flexible,Fazelzadeh}.

 \subsection{Translational dynamics}

 \begin{figure}[t]
    \centering
    \begin{tikzpicture}
            \draw (0,0) node[inner sep=0]{\includegraphics[width=0.9\linewidth ]{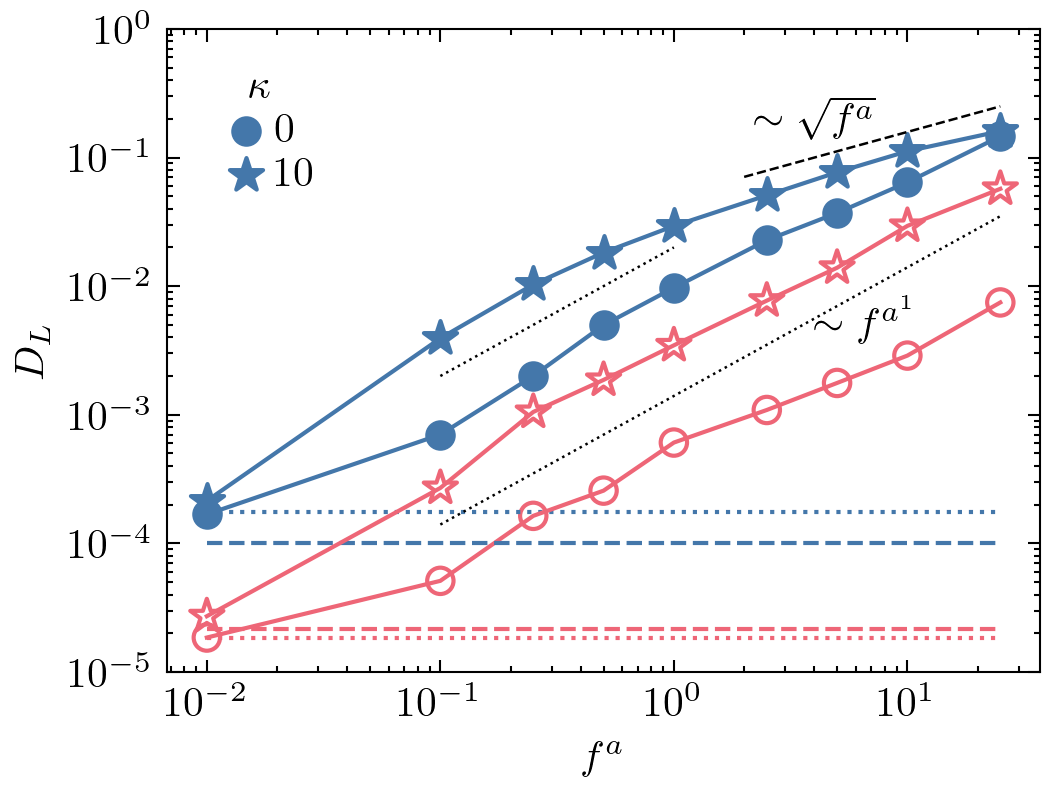}};
    \end{tikzpicture}
    \caption{\revise{The longtime diffusion coefficient $D_{L}$ versus active force $f^a$ for chains with $\kappa=0$ and $10$. Close and open symbols represent data for chains with and without hydrodynamic interaction, respectively. The black dotted and dashed lines show the scaling slopes of $f^{a^{1}}$ and \revise{$\sqrt{f^{a}}$}. The dotted and dashed lines show the values at $f^a=0$ for $\kappa=0$ and $10$, respectively.}}
    \label{fig:Dl}
\end{figure}

Next we explore the effects of HI on the translational dynamics of active polymers   by computing the mean squared displacement (MSD) of the center of mass. Defining the position of the center of mass at any time as $\vec{R}_{cm}(t)=\frac{1}{N_m}\sum \vec{R}_i(t)$, the MSD is calculated as $\langle \Delta \vec{R}^2_{cm}(t)\rangle=\langle|\vec{R}_{cm}(t)-\vec{R}_{cm}(0)|^2\rangle$. Fig.~\ref{fig:msd} presents MSD curves as functions of the lag time  for  fully flexible and semiflexible polymers, at two active forces $f^a=0.1$ and $25$ with and without HI. In all of the MSD plots, we observe three distinct scaling regimes. At very short lag times ($t<10^3$ for $f^a=0.1$ and $t<1$ for $f^a=25$), the fluctuations of random forces predominate the motion of the center of mass. Governed by thermal diffusion, the MSD curves obey the form $\langle \Delta \vec{R}^2_{cm}(t)\rangle=6D_{\text{Passive}}t$. For dry active polymers, $D_{\text{Passive}} (\kappa=0) \approx 2.4 \times 10^{-5}$ and $D_{\text{Passive}} (\kappa=10) \approx 2.7 \times 10^{-5}$, which agree well with  the theoretical prediction for the diffusion coefficient of a  passive polymer of chain length $N=50$ without HI, {\it i.e.}, $D_{\text{Passive}}=k_B T / (N\gamma)\approx 2.6\times10^{-5}$. For the wet active polymers the short-time passive diffusion 
  $D_{\text{Passive}}\approx 2\times 10^{-4}$ is one order of magnitude larger than their dry counterparts, reconfirming importance of HI on the dynamics even in the passive limit.

At intermediate timescales, the influence of activity  on MSD becomes visible. At this regime the MSD graphs are super-diffusive, \emph{i.e.}, $\langle \Delta \vec{R}^2_{cm}(t)\rangle \sim t^{\alpha}$ with $\alpha > 1$. At high active force of $f^a=25$, the intermediate regime is ballistic ($\alpha=2$), which means the motion of the center of mass of the chains can be roughly interpreted as moving straight with a constant speed in the direction of the sum of the total active force and backflow. As discussed earlier, the total active force  $\vec{F}^a(t)=f^a \vec{R}_{e}(t)/\ell $ and net backflow force  $\vec{F}^b_{cm}(t)= \sum _{i=1}^{N} \vec{F}^b_{mono}(t) = -\frac{NM}{mN + MN_m}\vec{F}^a$ are both proportional to the end-to-end vector. Therefore, we expect that the timescale at which the center of mass moves ballistically is set by the relaxation time of the end-to-end vector, \emph{i.e.}, $\tau_{e}$. At low active force of $f^a=0.1$, as we see in Fig.~\ref{fig:taue}, the orientational relaxation time is of the same order of the passive relaxation time.  Consequently, the thermal fluctuations have enough time to perturb the active directed motion of the polymer resulting in a superdiffusive MSD in the intermediate regime with $1<\alpha<2$.  

Finally at larger lag times when $t \gg\tau_e$ the end-to-end vector (hence the total self propulsion) becomes uncorrelated to its initial direction and a final diffusion regime with enhanced diffusion coefficient emerges. The MSD curves at this regime follow the form $\langle \Delta \vec{R}^2_{cm}(t\gg \tau_e)\rangle = 6D_L t$, where $D_L$ is the enhanced longtime diffusion coefficient. We extract the values of the diffusion coefficients by linear fits to the MSD curves at large lag times. The values of $D_L$ against activity are presented in Fig.~\ref{fig:Dl} for wet and dry active polymers with bending stiffness values of $\kappa=0$ and $10$.  
In general, the wet chains exhibit more enhanced longtime dynamics.  For dry active polymers, the $D_L$ increases linearly with activity in agreement with theoretical predictions~\cite{Fazelzadeh,Winkler-Theoretical-LinearPol}. Interestingly for flexible wet chains with $\kappa=0$,  even with incorporation of HI, the $D_L$ increases linearly with activity.  However for semiflexible wet active polymers, while at lower activity levels ($f^a<1$) the scaling behaviour of $D_L$ is linear in activity, at higher levels of activity where $f^a\geq 2.5$, another scaling  regime emerges where   \revise{$D_L\sim \sqrt{f^{a}}$}. The emergence of a new scaling regime at high active forces can be attributed to the HI-induced dramatic shrinkage of semiflexible wet chains, as visible from Fig.~\ref{fig:Re}. Since the total active force is proportional to the end-to-end distance, the steep drop in $\sqrt{\langle R_e^2 \rangle}$ decreases the scaling exponent of $D_L$ with respect to $f^a$.

 \section{Conclusions \label{Sec:conclusion}}
In this paper, we presented a comparative study of the conformational and dynamical features of single tangentially-driven polymer chains in dilute solutions with and without hydrodynamic interactions (HI). We performed computer simulations using coarse-grained bead-spring models, incorporating fluid-mediated interactions using the MPCD method, revealing significant role of HI on conformational and dynamical properties of tangentially-driven active polymers. 

Studying the conformational  features of  wet flexible and  semiflexible active polymers, we find that HI lead to shrinkage of active  polymers, the extent of which  increases with the activity level. The most remarkable shrinkage occurs for semiflexible polymers where HI induce a transition from an extended polymer conformation to a helix-like conformations. This remarkable shrinkage results in the decrease of effective persistence length of semiflexible polymers which arises from combined effects of activity and HI.  Our findings highlight the significant conformational changes driven by the interplay of activity, elasticity, and hydrodynamic interactions (HI). This contrasts with dilute solutions of passive polymers, where conformational features are only weakly affected by HI~\cite{passive_wet_polymer}.

 Exploring the dynamics of active polymers, we find that both translational and rotational dynamics of wet active polymers are accelerated compared to their dry counterparts.  Interestingly, the orientational dynamics of active polymers does not depend on their degree of flexibility but the activity level. HI decrease the orientational relaxation time  of wet active polymers by an order of magnitude, irrespective of the activity level. Likewise, HI increase the center of mass longtime diffusion of wet polymers at low activity levels by an order of magnitude. However, the HI-induced enhancement of longtime diffusion is mitigated at larger activity levels, particularly for semiflexible polymers due to their shrinkage. This trend can be understood given that the active contribution to $D_L$  becoming dominant at large active forces is  proportional to $\langle R_e^2\rangle f^a$. As orientational relaxation time is independent of bending stiffness and the end-to-end distance of wet flexible and semiflexible polymers at large activity levels becomes very similar, their $D_L$ values also approach each other.

In conclusion, our work offers novel insights into the interplay between activity and fluid-mediated interactions on the conformation and dynamics of tangentially-driven active polymers for which active force and conformation are coupled to each other. By implementing active chain hydrodynamics using the MCPD  fluid method, we pave the way for future investigations into the dynamics of tangentially-driven polymers in complex environments, such as confinement and crowded media.


\begin{acknowledgements}
We acknowledge E. Irani and R. Winkler for fruitful discussions and M. Howard for the advice to make home-made modification of Hoomd-blue to couple the active polymer dynamics to the MPCD fluid. The computations were carried out on the Dutch National e-Infrastructure with the support of the SURF Cooperative. 
This work was part of the D-ITP consortium, a program of the Netherlands Organization for Scientific Research (NWO) that is funded by the Dutch Ministry of Education, Culture and Science (OCW). 
\end{acknowledgements}

\begin{figure}[t]
    \centering
    \begin{tikzpicture}
            \draw (0,0) node[inner sep=0]{\includegraphics[width=1\linewidth ]{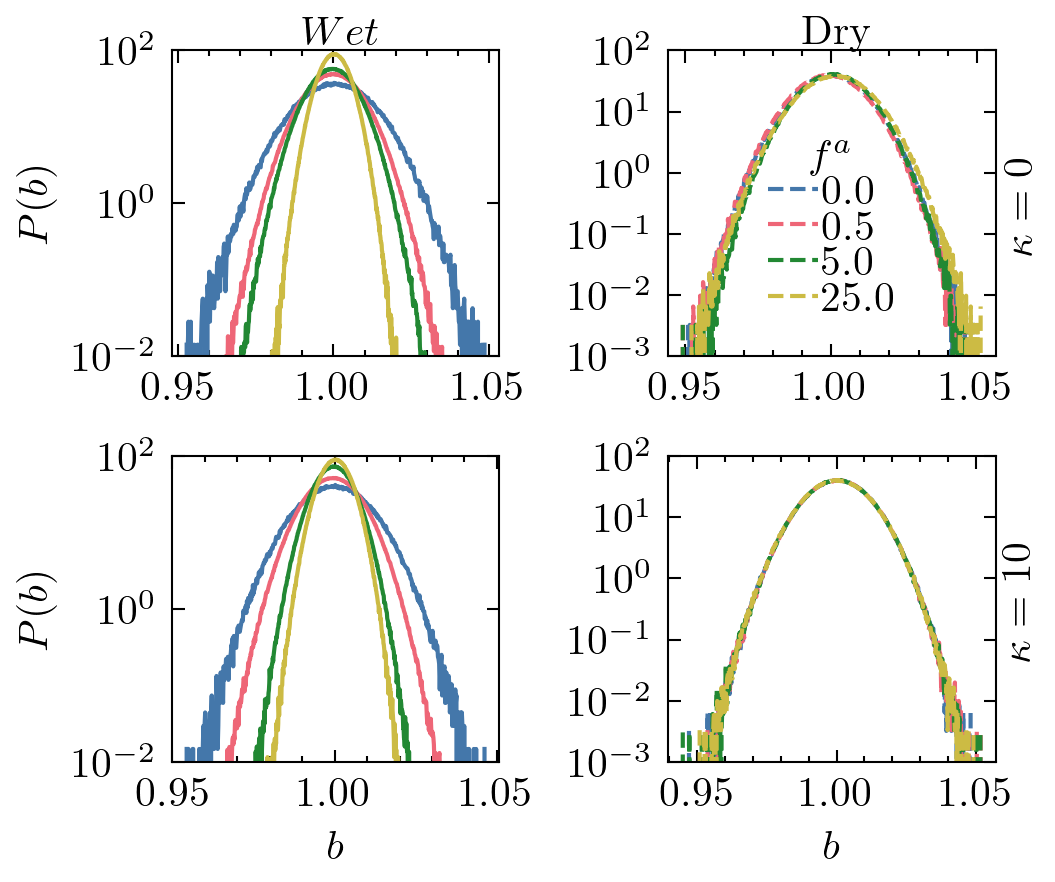}};
            \draw {(-2.6,2.9) node{\textbf{(a)}}};
            \draw {(1.5,2.9) node{\textbf{(b)}}};
            \draw {(-2.6,-0.4) node{\textbf{(c)}}};
            \draw {(1.5,-0.4) node{\textbf{(d)}}};
    \end{tikzpicture}
    \caption{ \revise{The bond length distribution $P(b)$ for wet and dry polymers with bending stiffness of $\kappa=0$ and $10$. }}
    \label{fig:bonds}
\end{figure}

\appendix
\revise{\section{Distribution of bond length}
The distribution of bond lengths are presented in Fig.\ref{fig:bonds} for wet and dry chains with $\kappa=0$ and $10$ for several activity values. We have increased the value of $k_s$ proportionally to active force in order to prevent over-stretching or compression of bonds and keeping a fixed mean bond length equal to rest bond length $ \langle b \rangle=1$. The distribution of bond length  $b$  for flexible and semiflexible polymers at low and high activity is shown in Fig.~\ref{fig:bonds} (a-d). As can be seen the width  of distribution is less than $\%5$.}

\revise{\section{Simulation with larger box size} \label{APP:L100}
In order to check for finite box size effects and any potential  self-interaction of active polymers due to long-range hydrodynamic interactions, we ran simulations with a box size of $L_{\text{box}} = 100$ for semiflexible polymers ($\kappa = 10$) at activity levels $f^a = 0.1$ and $25$. To compare polymer conformations between the larger box and the original box ($L_{\text{box}} = 60$), we computed $\sqrt{\langle R_e^2 \rangle}$. For $L_{\text{box}} = 100$, the values are 24.8 and 7.2, which closely match the values of 25.0 and 7.2 for $L_{\text{box}} = 60$. Similarly, for polymer dynamics, we calculated $D_L$. For $L_{\text{box}} = 100$, we found $D_L = 3.8 \times 10^{-3}$ and $1.7 \times 10^{-1}$, which are nearly identical to $3.9 \times 10^{-3}$ and $1.7 \times 10^{-1}$ for $L_{\text{box}} = 60$.
 }
\begin{figure}[ht]
    \centering
    \begin{tikzpicture}
            \draw (0,0) node[inner sep=0]{\includegraphics[width=1\linewidth ]{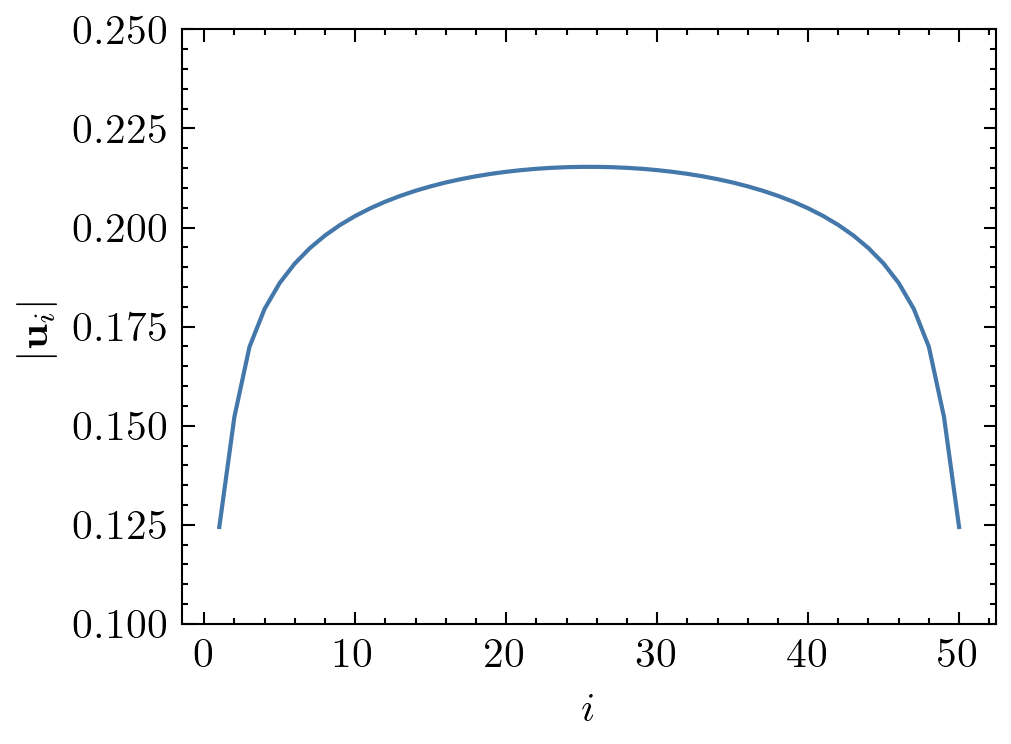}};
    \end{tikzpicture}
    \caption{ \revise{The distribution of the flow speed at positions of monomers of the straight polymer with $\kappa=10$ and $f^a=25$. }}
    \label{fig:oseen}
\end{figure}
\revise{\section{Evaluation of the flow field of straight active polymer}

We consider a polymer in an initial straight conformation where the distance between adjacent monomers equals the equilibrium bond length. In this configuration, the only forces acting on each monomer are the active forces. The flow field generated by the active forces can be calculated using the Oseen tensor~\cite{oseen1910}. The position of bead $j$ in the straight polymer is $\vec{R}_j = (j-1)\ell\hat{x}$ for $j = 1, 2, \dots, N$. The tangential active forces are $\vec{f}^a_j=f^a\hat{x}$ for $2 \leq j \leq 49$ and $\vec{f}^a_j=f^a\hat{x}/2$ for $j = 1$ and $N$. 

The resulted flow field at the vicinity of monomer $i$ is then given by the sum of flow fields created by all other active forces at the position of monomers $j$:
\begin{equation}
    \vec{u}(\vec{R}_i)=\sum_{j\neq i} \Bar{\Bar{G}}(\Vec{R}_{ij}).\vec{f}^a_j +  \vec{f}^a_i/\gamma \;,
     \label{eq:uoseen}
\end{equation}
 where  $\Bar{\Bar{G}}(\Vec{R}_{ij}) = \dfrac{1}{8 \pi \eta R_{ij}}(1 + \dfrac{\Vec{R}_{ij}\otimes\Vec{R}_{ij}}{R_{ij}^2} )$ is the Oseen tensor,  $\vec{R}_{ij}=\Vec{R}_i - \Vec{R}_j $ and   the second term on the RHS of Eq.~\eqref{eq:uoseen} corresponds to the flow field of the monomer $i$ at its own position. Using the definitions of $\vec{R}_j$ and $\vec{R}_i$, we find:
\begin{align}
    & \vec{u}_i := \vec{u}(\vec{R}_i) = \vec{f}^a_i/\gamma + \sum_{j\neq i} \dfrac{\vec{f}^a_j}{4 \pi \eta |i-j|\ell} \;,
\end{align}
where $\vec{u}_i$ is parallel to $\hat{x}$, inferred from the symmetry of the straight conformation and the tangential nature of the force monopoles along $\hat{x}$. We evaluate the flow speed $|\vec{u}_i|$ along the chain for $N=50$ as shown in Fig.~\ref{fig:oseen}. The flow speed is heterogeneous, peaking in the middle and decreasing towards the ends. The faster middle segment pulls the tail, helping maintain the straight conformation for $i < 25$. However, the collision between the fast middle part and slow frontal region causes the chain to buckle at the front.
}


 \bibliography{active_polymer.bib}
   \end{document}